\documentclass[accepted=2018-06-27]{quantumarticle}
\usepackage{graphicx}
\usepackage{hyperref}
\usepackage{verbatim}
\usepackage{bbold,bbm}
\usepackage{amsmath,amssymb}

\begin{document}

\title{Hamiltonians for one-way quantum repeaters}
\author{Filippo M. Miatto}
\affiliation{Institute for Quantum Computing and Department of Physics and Astronomy, University of Waterloo, 200 University Ave.~W, N2L 3G1 Waterloo, Ontario, Canada}
\affiliation{T\'el\'ecom ParisTech, LTCI, Universit\'e Paris Saclay, 46 Rue Barrault, 75013 Paris, France}
\orcid{0000-0002-6684-8341}
\author{Michael Epping}
\affiliation{Institute for Quantum Computing and Department of Physics and Astronomy, University of Waterloo, 200 University Ave.~W, N2L 3G1 Waterloo, Ontario, Canada}
\author{Norbert L\"utkenhaus}
\affiliation{Institute for Quantum Computing and Department of Physics and Astronomy, University of Waterloo, 200 University Ave.~W, N2L 3G1 Waterloo, Ontario, Canada}
\orcid{0000-0002-4897-3376}

\begin{abstract}
Quantum information degrades over distance due to the unavoidable imperfections of the transmission channels, with loss as the leading factor. This simple fact hinders quantum communication, as it relies on propagating quantum systems. A solution to this issue is to introduce quantum repeaters at regular intervals along a lossy channel, to revive the quantum signal. In this work we study unitary one-way quantum repeaters, which do not need to perform measurements and do not require quantum memories, and are therefore considerably simpler than other schemes. We introduce and analyze two methods to construct Hamiltonians that generate a repeater interaction that can beat the fundamental repeaterless key rate bound even in the presence of an additional coupling loss, with signals that contain only a handful of photons. 
The natural evolution of this work will be to approximate a repeater interaction by combining simple optical elements.
\end{abstract}
\maketitle

\section{Introduction}
Quantum physics plays an increasingly important role in the development of future technologies. The transmission of information is an example of this trend, where the benefit of exploiting quantum effects has motivated several ideas, such as quantum money, quantum key distribution (QKD), quantum fingerprinting, quantum digital signatures, and several others \cite{bennett1983quantum,bb84,buhrman2001quantum,gottesman2001quantum,gisin2002quantum,lo2014secure}. 
Eventually, we will build a quantum internet: a communication infrastructure that will enable distributed quantum information tasks on a global scale \cite{kimble2008quantum,lloyd2004infrastructure}.

It would be very convenient if we could rely as much as possible on the current infrastructure, such as the optical fiber network, to propagate quantum states of light as the main carrier of quantum information. Light is cheap, fast, and it does not require vacuum or low temperatures. However, at the present moment, we cannot send photons reliably to distant parties: optical fibers are ``leaky'' and deteriorate the quantum properties of the signal. The fact that quantum information cannot travel far enough constitutes one of the outstanding challenges in the field \cite{sangouard2011quantum}, and it is the main focus of our work.

Note that as optical modes are like harmonic oscillators, photon loss is equivalent to the relaxation of a harmonic oscillator towards  equilibrium \cite{loudon2000quantum}. This makes our findings applicable to any domain in which one needs to fight relaxation (e.g. fields in a cavity, nanoscale resonators, etc). 

\section{Preliminaries}
We know from the theory of quantum information that we can ``stretch'' the distance between two entangled systems by a process that involves entanglement swapping, which is a core component of the first generation of quantum repeaters \cite{briegel1998quantum}. 
The newest generation of repeater architectures relies instead on quantum error correction, which allows for ``one-way'' functionality by removing the need to broadcast measurement results backwards and forwards \cite{knill1996concatenated,munro2012quantum, muralidharan2014ultrafast,muralidharan2016optimal}. It is even possible to omit measurements altogether: just as one can implement quantum error correction by measuring the error syndrome and then applying the appropriate correction, it is also possible to do so coherently by using an ancillary system initialized in a fiducial state $|0\rangle$ to store the index $i$ of each syndrome and of the corresponding Kraus operator:
\begin{align}
\label{urep}
U_\mathrm{rep}=\sum_{i=1}^{\mathrm{syndromes}}R^{(i)}\otimes |i\rangle\langle0|.
\end{align}
In this expression, $R^{(i)} = U_iP^{(i)}$ where $P^{(i)}$ is the projector onto the $i$-th syndrome space, and $U_i$ is the unitary that brings the system from the $i$-th syndrome space back into codespace.
Note that expression \eqref{urep} is intended as an isometry. The need for an ancilla is a simple consequence of the Stinespring dilation theorem \cite{nielsen2002quantum}, according to which any CPTP channel can be described through a unitary interaction with a larger system.

In this work we present two recipes to build families of Hamiltonians that generate such unitary one-way quantum repeaters. 

In order to evaluate the performance of our repeaters, we will compare the entanglement between two systems (one of which is travelling in a lossy channel interspersed with repeaters), to the repeaterless key rate bound \cite{takeoka2015fundamental,pirandola2017fundamental}, see also \cite{wilde2017converse} for results in the non-asymptotic regime. In particular, \cite{pirandola2017fundamental} found that the tightest upper bound of the QKD rate per optical mode that can be achieved using only a lossy channel with transmissivity $\eta$ is
equal to:
\begin{align}
R(\eta) = \log_2\frac{1}{1-\eta}.
\end{align}
What we found is that we can beat this repeaterless bound even if each repeater introduces a mild coupling loss between the optical fibre and the repeater device. Furthermore, the optimized separation between the repeater stations turns out to be rather reasonable, on the order of a few kilometers.

The complexity of our repeater Hamiltonians depends on the code space that they protect and on the fiducial ancilla states. Consequently, if we pick them wisely, we have the potential to obtain \emph{simple} Hamiltonians that  can be implemented by combining a few known optical elements, such as single photon sources, beamsplitters, squeezers, four wave mixers, etc.
Such a quest for simplification is outside the scope of the present work, but it is part of our ongoing research.

\section{Working principles of a one-way quantum repeater}
Our task is to protect the quantum information that is encoded in a photonic system. Although in this particular context we lack the typical qubit structure of QEC codes, if we treat photon loss as an ``error'', we can still apply QEC successfully to our problem. Note that the set of correctable errors cannot include all of the possible loss operators (e.g. including the loss of all photons), but rather a few of them. How many depends on the particular code in question. 

Furthermore, in practice the correction process is not going to be perfect and it can itself introduce errors and noise. This enforces an optimal separation between the repeaters, as well as other important considerations regarding the noise parameters. In this work we describe an ideal repeater, i.e. one that does not introduce any additional noise. In this we consider the method of exact error correction \cite{shor1995scheme, knill1997theory} to illustrate our approach of Hamiltonian Quantum Repeaters. It will be interesting to extend this line of consideration also to approximate error correction \cite{tyson2010two, beny2010general, ng2010simple, mandayam2012towards} though we leave that to upcoming work. For this reason it is sufficient for us to consider exact  and not approximate quantum error correction.

In all that follows, we will seek the protection of a two-dimensional subspace within each signal from the loss of a single photon. If this is done frequently enough, and if we separate the repeaters such that the single-photon loss is the dominant error, we can witness the benefits of a repeater action. All that we need from QEC is encapsulated in the QEC condition in the form $P E^\dagger_i E_j P =\delta_{ij} c_{i}P$ where $c_i$ is the probability of the $i$-th error, $P$ is the projector onto the protected subspace (called the ``code space'') and \{$E_i$\} are the correctable effect operators of the undesired interaction, which map $P$ to the orthogonal spaces $P^{(i)}=E_i P E_i^\dagger/\mathrm{\frac{1}{2}Tr}(E_i P E_i^\dagger)$ \cite{nielsen2002quantum}. The fact that such spaces are orthogonal to each other and to $P$ means that the quantum information has not left the system, it is just in a different subspace and it can be brought back. 

To understand what the error operators are, we need a description of the lossy channel. There are three main ways to describe a single-mode lossy channel of transmissivity $\eta$:

(i) As a quantum channel in Kraus form (also e.g. in \cite{fan2008new}):
\begin{align}
\mathcal{L}_\eta[\rho]=\sum_{k=0}^\infty\frac{(1-\eta)^k}{k!}\sqrt{\eta}^{N}{ a}^{k} \rho { a^\dagger}^{k}\sqrt{\eta}^{ N},
\label{kraus}
\end{align}
where $N=a^\dagger a$ is the number operator and $a$ is the annihilation operator. The Kraus operators are therefore $A_k = \frac{(1-\eta)^{k/2}}{\sqrt{k!}}\sqrt{\eta}^N a^k$.

(ii) As the action of a beamsplitter in each mode (effectively the Stinespring representation of the channel):
\begin{align}
\mathcal{L}_\eta[\rho]=\mathrm{Tr}_B\left[U_{AB} (\rho_A\otimes |0\rangle\langle 0|_B) U_{AB}^\dagger\right],
\label{stinespring}
\end{align}
where $U_{AB}=\exp[i\phi(a^\dagger b+ab^\dagger)]$ is a beamsplitter between the signal and a local environment in the vacuum state.

(iii) Through the master equation \cite{loudon2000quantum}:
\begin{align}
\dot\rho = \frac{\gamma}{2}\left(2a\rho a^\dagger-\{N,\rho\} \right).
\label{master}
\end{align}
The connection between the three representations is the relation $\eta= \cos(\phi)^2=\exp(-\gamma t)$. This channel generalizes to an $M$-modes lossy channel as a simple tensor product of $M$ channels. We will indicate the Kraus operator corresponding to losing $k$ photons from mode $i$ and none from the other modes as $A_k^{(i)}=\frac{(1-\eta)^{k/2}}{\sqrt{k!}}\sqrt{\eta}^N a_i^{k}$ (where $N=\sum_i a_i^\dagger a_i$). As we concentrate on correcting single photon losses, we need the single-photon loss operator in the $i$-th mode: $E_i:=A_1^{(i)}=\sqrt{1-\eta}\sqrt{\eta}^{ N}{ a_i}$. If we replace these operators into the error-correction conditions and take the trace of both sides, we obtain that the probability of the $i$-th error ($i \geq 1$) is:
\begin{align}
\label{cj}
c_{i} = \frac{1-\eta}{d} \mathrm{Tr}\left(Pa_i^\dagger a_i\eta^{N-1}\right)
\end{align}
where we used the operator identity $f(N)a = af(N-1)$ and $d$ is the dimension of the protected subspace, which comes from $\mathrm{Tr}(P) = d$. In our case, as we are protecting a two-dimensional subspace, $d=2$.
Eq.~\eqref{cj} means that, understandably, the more photons a code uses the easier it is to lose one, which gives rise to an interesting trade-off between the probability to lose a photon and the ability to protect from photon loss. From inspection, it appears that a good choice is to make sure we can correct for a single loss in each mode using as few photons as possible in the code. 

Note that in case of no loss, we have a non-trivial evolution: the no-loss operator is $E_0 = A_0^{(i)} = \sqrt{\eta}^N$, which is not the identity operator. This means that if the code space $P$ is not an eigenspace of the total photon number operator, it will evolve non-trivially even if no losses occur. 

As anticipated, a unitary operation is sufficient to rotate $P^{(i)}$ back to $P$, but there is a catch: the operation that we should apply to the system depends on the error syndrome. As we have more than one subspace that should rotate back to $P$,  this is not a unitary operation unless we introduce an ancillary system, as in Eq.~\eqref{urep}. So the quantum repeater must produce a fresh local ancilla, it must have the incoming system interact with it, and then it can discard the used ancilla. The output of the repeater is the recovered system (unless more losses occurred than the code can handle). Note that in principle we could measure the used ancilla to detect the syndrome, but the repeater would work equally well if we did not.

To summarize: we choose to put our quantum information in protected subspaces that do not become entangled with the environment upon photon loss, but rather they rotate to an orthogonal subspace. This enables us to rotate them back without involving measurements if we supply this operation with a `catalytic' ancilla which we can discard at the end of the operation.

\section{Repeater Hamiltonians}
In this section we present two families of Hamiltonians that generate a repeater unitary. Each Hamiltonian will depend on the code space $P$ as well as the initial ancilla state, so one can view them as a \emph{resource}, which can be chosen wisely to yield simpler Hamiltonians in the sense mentioned in the introduction. Note that a well-designed repeater action will not let the uncorrectable error spaces to eventually couple to code space (for instance after a series of several losses and repeaters). So the action of $U_\mathrm{rep}$ on spaces that will never interfere with code space can be thought of as another resource, as we can use the freedom to choose this otherwise irrelevant action to simplify the Hamiltonian.
We present both architectures because although they are related by a system-ancilla swap after the repeater, the Hamiltonians are rather different.

\subsection{Direct unitary}
\begin{figure}[h!]
\centering
\includegraphics[width=0.9\columnwidth]{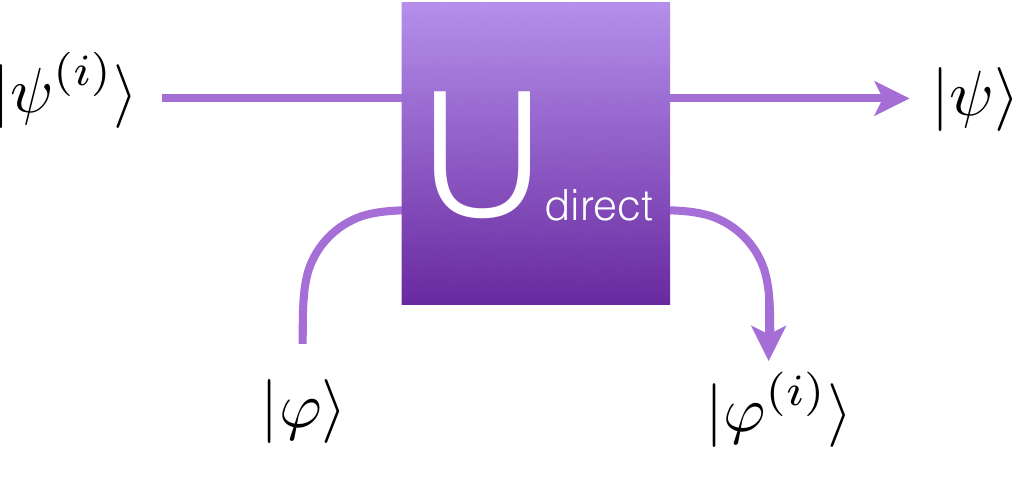}
\caption{\label{direct}In the direct architecture, the system interacts with a local ancilla, it ``transfers the error'' and then proceeds onwards. Measuring the used ancilla only gives us information about the loss, which does not influence the performance of the repeater.}
\end{figure}
The first architecture (Fig.~\ref{direct}) that we introduce is that of a repeater which lets the input system ``transfer'' the loss errors to the ancilla. Once this operation is complete, we have essentially recovered the initial state. In this construction the initial state of the ancilla is free to live in a $K$-dimensional space, and one is free to pick any value for $K\geq 1$. The Hamiltonian is
\begin{align}
H_\mathrm{direct}=\sum_{k=1}^K\sum_{i=1}^M\left(\Psi_{0k}^{(i)}+\Psi_{1k}^{(i)}\right),
\end{align}
where $M$ is the number of modes, $\Psi_{jk}^{(i)}$ is a projector onto $|\Psi_{jk}^{(i)}\rangle = \frac{1}{\sqrt{2}}(|j\rangle_s|k^{(i)}\rangle_a-|j^{(i)}\rangle_s|k\rangle_a$, and the subscripts $s$ and $a$ indicate the system and the ancilla. Here we must comply with the error spaces that are imposed by the loss operators, i.e. for the system we have $|j^{(i)}\rangle=E_i|j\rangle/\sqrt{\langle j|E_i^\dagger E_i|j\rangle}$, however we have complete freedom (even in terms of number of modes) to pick the ancilla states $|k\rangle_a$ and $|k^{(i)}\rangle_a$ as long as all together they form an orthonormal set. As this Hamiltonian acts as the identity operator on the space of non-correctable errors (two or more photon lost), it automatically avoids these error spaces to couple back to the code space. It is possible to simplify the direct Hamiltonian by minimizing the dimension of the initial ancilla space (i.e. $K=1$):
\begin{align}
H_\mathrm{direct}'=\sum_{i=1}^M\left(\Psi_{0}^{(i)}+\Psi_{1}^{(i)}\right),
\end{align}
where $\Psi_{j}^{(i)}$ projects onto $|\Psi_{j}^{(i)}\rangle = \frac{1}{\sqrt{2}}(|j\rangle_s|0^{(i)}\rangle_a-|j^{(i)}\rangle_s|0\rangle_a)$.
Note that in the construction of $H_\mathrm{direct}$ (as opposed to the construction of $H_\mathrm{swap}$ in the next subsection) we excluded the no-loss ($i=0$) element, as the repeater should act as the identity if no error occurs. We remark that in case a code had a non-trivial evolution under zero loss, one might  need to include a term to act in this eventuality.

The fact that the Hamiltonian is a projector simplifies enormously the computation of the unitary, which implements the correction at time $t = \pi$:
\begin{align}
U_\textrm{direct}=\exp\left(i\pi H_\textrm{direct}\right)=\mathbbm{1}-2H_\mathrm{direct}.
\end{align}
It is a simple computation to verify that $U_\mathrm{direct}$ indeed corrects an error by ``transferring'' it to the ancilla:
\begin{align}
\label{directU}
U_\mathrm{direct}|\psi^{(i)}\rangle_s\otimes|\varphi\rangle_a = |\psi\rangle_s\otimes|\varphi^{(i)}\rangle_a,
\end{align}
where $|\psi^{(i)}\rangle_s=\alpha|0^{(i)}\rangle_s+\beta|1^{(i)}\rangle_s$ is the state of the system after the error $i$, and $|\varphi\rangle_a=\sum_k\gamma_k|k\rangle_a$ is the initial ancilla state ($|\varphi\rangle_a$ is fixed, for example the vacuum).

\subsection{SWAP unitary}
\begin{figure}[h!]
\centering
\includegraphics[width=0.9\columnwidth]{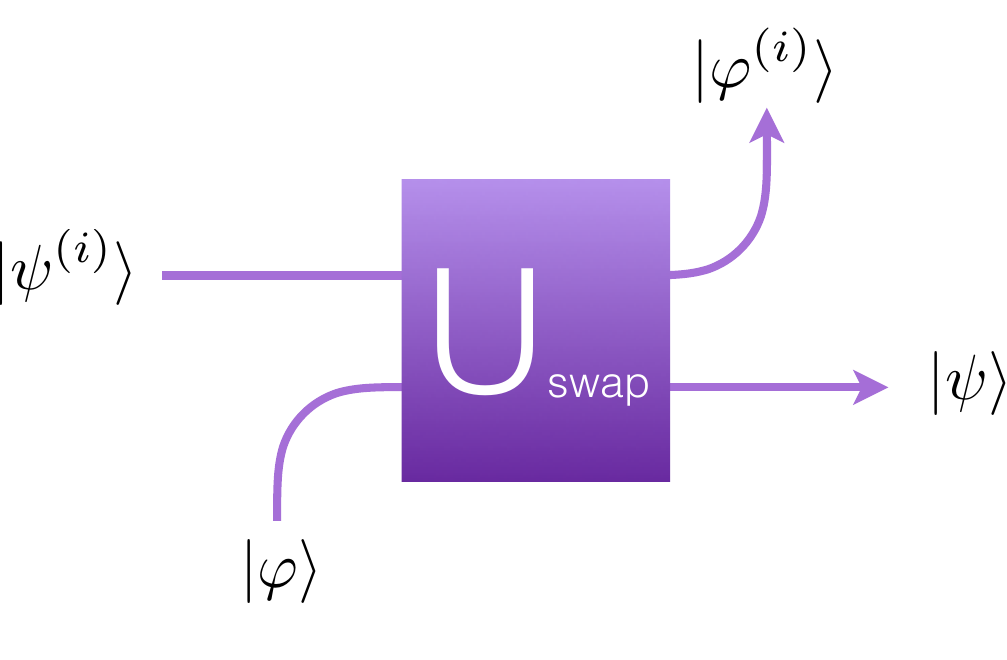}
\caption{\label{swap}In the ``swap'' repeater configuration, the system and the ancilla remain in their respective spaces (the error space $P^{(i)}$ and code space), but they swap states: the system acquires the state of the ancilla and the ancilla acquires the state of the system and then proceeds onwards.}
\end{figure}

The second architecture (Fig.~\ref{swap}) is a repeater that ``swaps'' the state of the input system with the state of an ancilla prepared in code space, while leaving the errors behind. More specifically, the SWAP operation occurs between $P$ (where the ancilla needs to be initialized) and the space where the system currently lives, which is $P$ if no error occurred and it is $P^{(i)}$ if a single photon loss from the $i$-th mode occurred.
The Hamiltonian of this repeater is:
\begin{align}
H_\mathrm{swap}&=\sum_{i=0}^M\Phi^{(i)}
\end{align}
where $\Phi^{(i)}$ projects onto $|\Phi^{(i)}\rangle=\frac{1}{\sqrt{2}}(|0^{(i)}\rangle_s|1\rangle_a-|1^{(i)}\rangle_s|0\rangle_a)$. Similar to $H_{direct}$, this Hamiltonian also automatically avoids the uncorrectable error spaces to couple back to the code space. Note that now in case of no error the repeater still needs to swap, so we have to include the $i=0$ case.  The correction unitary is again given by
\begin{align}
U_\textrm{swap}=\exp\left(i \pi H_\textrm{swap}\right).
\end{align}
And it is again a simple computation to verify that the unitary corrects by swapping (compare with Eq.~\eqref{directU}):
\begin{align}
U_\mathrm{swap}|\psi^{(i)}\rangle_s\otimes|\varphi\rangle_a = |\varphi^{(i)}\rangle_s\otimes|\psi\rangle_a,
\end{align}
where $|\psi^{(i)}\rangle_s=\alpha|0^{(i)}\rangle_s+\beta|1^{(i)}\rangle_s$ is the state of the system after the error $i$, and $|\varphi\rangle_a=\gamma_0|0\rangle_a+\gamma_1|1\rangle_a$ is the initial ancilla state in code space.

\section{Beating the repeaterless bound}
In this section we evaluate how well our repeaters protect the quantum information along a lossy channel. Our goal is to beat the repeaterless bound asymptotically, which in terms of distance $x$ with $\alpha$ dB/Km of loss is $R(x) = -\log_2(1-10^{-\alpha x/10})$. For a telecom fibre, $\alpha\approx0.2$ dB/Km.

We present a single-mode code to show the effect of the uncorrected no-loss operator $\sqrt\eta^N$ and we will compare a four-photon two-mode code \cite{chuang1997bosonic} with a three-photon three-mode code \cite{wasilewski2007protecting} to show the advantage of using fewer photons. We supply the logical codewords that span $P$ explicitly in terms of the number states in Fock space:
\begin{enumerate}
\item \emph{single-mode code}: Logical states $|0_L\rangle = |1\rangle$ and $|1_L\rangle = |3\rangle$.
\item \emph{two-mode code}: Logical states $|0_L\rangle = \frac{|4,0\rangle+|0,4\rangle}{\sqrt{2}}$ and $|1_L\rangle = |2,2\rangle $.
\item \emph{three-mode code}: Logical states $|0_L\rangle = |1,1,1\rangle$ and $|1_L\rangle = \frac{|0,0,3\rangle+|0,3,0\rangle+|3,0,0\rangle}{\sqrt 3}$.
\end{enumerate}
These are just simple didactical examples, for an excellent in-depth presentation of bosonic codes see \cite{michael2016new, li2017cat, albert2017performance}.

We initialize an entangled state $|\psi\rangle_{AB}=\frac{1}{\sqrt{2}}(|0,0_L\rangle+|1,1_L\rangle)$ using the logical states in one of the example codes presented above, and we propagate the $B$ system through a lossy channel with repeaters separated by a distance $L$. For codes that are eigenstates of the total photon number, after one segment comprising loss and repeater action, we find the original state in code space with probability $p_s$ (defined below). Since the uncorrectable error spaces do not couple back to the code space, we can extract $p_s^n$ secret bits of key after $n$ segments, which means that the condition to beat the repeaterless bound asymptotically is that the logarithmic slope of the key rate be larger than the logarithmic slope of the repeaterless bound: $p_s > \eta_s$, where $\eta_s$ is the transmissivity of a segment. The same condition also appears from different arguments, see \cite{namiki2016role, michael}.

\begin{figure}[h!]
\centering
\includegraphics[width=0.9\columnwidth]{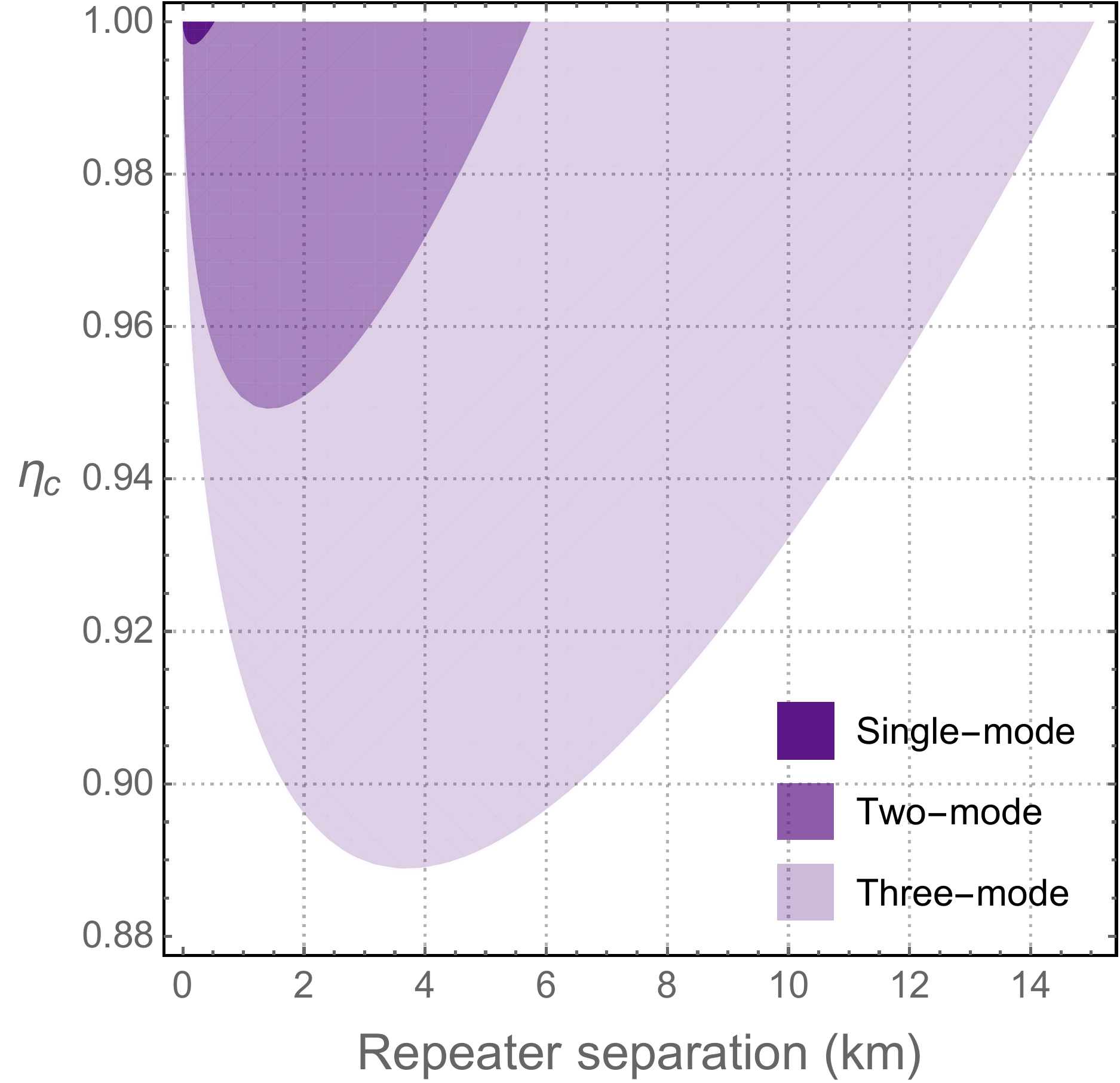}
\caption{\label{asymptotic}The shaded regions indicate for which coupling efficiency $\eta_c$ and repeater separation $L$ we can extract more key than the repeaterless bound allows (non-asymptotically for the single-mode code and asymptotically for the two- and the three-mode code). Notably, the three-mode code can withstand up to about 11\% of additional coupling loss between a fibre and a repeater station.}
\end{figure}

We also investigate the stability if we have coupling losses between an optical fibre and a repeater station. To do this, we consider the total loss of a segment to be composed of two parts: $\eta=\eta_c\eta_s$, where $\eta_c$ is the coupling efficiency of the fibre into the repeater and $\eta_s$ is the regular loss of the channel.
As the success probability $p_s$ depends on the code space $P$, on the repeater separation $L$ and on the coupling efficiency $\eta_c$, we can find for which values of these quantities we beat the repeaterless bound (see Fig.~\ref{asymptotic}). For the two-mode code, we have $p_s= \eta^4+4(1-\eta)\eta^3$, for the three-mode code we have $p_s = \eta^3+3(1-\eta)\eta^2$, where $\eta=\eta_c\eta_s$. Here we assume that a repeater does not couple the error spaces with the code space in undesired ways (e.g. by adding more photons than necessary, which may lead to the system entering back into code space upon further loss).

For codes that are not so well-behaved (such as the single-mode code in the examples), things are more complicated as the code spaces become distorted. In the case of our single-mode code, we need to compute explicitly the action of the lossy channel $\mathcal L_\eta$ from Eq.~\eqref{kraus} (where the total loss $\eta=\eta_s\eta_c$ is composed of the segment loss and the coupling loss) followed by the channel $\mathcal R$ that maps the signal from the input to the output of the repeater, which we can describe with $M+1$ Kraus operators: $R^{(0)} = (\mathbbm{1}-\sum_i P^{(i)})$ and $R^{(i)} = |0\rangle\langle0^{(i)}|+|1\rangle\langle1^{(i)}|$ for $1\leq i\leq M$ (see Eq.~\eqref{urep}). Note that we do not correct the distortion on the zero-photon loss (for some codes it can be possible, especially for small $\eta$ \cite{michael2016new}). We find:
\begin{align}
\mathcal{R}\circ\mathcal{L}_\eta|1\rangle\langle1| &= |1\rangle\langle1|\nonumber\\
\mathcal{R}\circ\mathcal{L}_\eta|1\rangle\langle3| &= (\eta^2+(1-\eta)\sqrt{3}\eta)|1\rangle\langle3|\nonumber\\
\mathcal{R}\circ\mathcal{L}_\eta|3\rangle\langle1| &= (\eta^2+(1-\eta)\sqrt{3}\eta)|3\rangle\langle1|\nonumber\\
\mathcal{R}\circ\mathcal{L}_\eta|3\rangle\langle3| &= (3\eta^2-2\eta^3)|3\rangle\langle3|+(1-3\eta^2+2\eta^3)|1\rangle\langle1|\nonumber
\end{align}
This mapping allows us to compute the $n$-fold application of $\mathcal{R}\circ\mathcal{L}_\eta$. After $n$ segments, the state is
\begin{align}
\psi_{AB}^{(n)}&=\frac{1}{2}|0,1\rangle\langle0,1|+\frac{1}{2}(1-(3\eta^2+2\eta^3)^n)|1,1\rangle\langle1,1|\nonumber\\
&\frac{1}{2}(\eta^2+(1-\eta)\sqrt{3}\eta)^n(|0,1\rangle\langle1,3|+|1,3\rangle\langle0,1|)\nonumber\\&+\frac{1}{2}(3\eta^2-2\eta^3)^n|1,3\rangle\langle1,3|
\end{align}

If we perform the six-states protocol \cite{bruss1998optimal,lo2001proof} with this state, we can compute the key rate by following the results in appendix A of \cite{scarani2009security}. We resort to the six-state protocol because the distortion causes errors, and thus the simple consideration of success probabilities will not be sufficient to guarantee the cross-over.
What we find is that the region of parameter space that allows us to beat the repeaterless bound \emph{at some distance} (note: not asymptotically) is very small, but it exists (see Fig.~\ref{asymptotic}). We remark that it might be possible to increase the single-mode key rate by other means (correcting the zero-loss effect, noisy pre-processing, two-way codes, etc...).

We can finally plot for each code the parameter region that allows us to beat the repeaterless bound (non-asymptotically for the single-mode code and asymptotically for the two- and three-mode codes) in a telecom fiber (Fig.~\ref{asymptotic}).

\section{Considerations}
\begin{figure}[h!]
\centering
\includegraphics[width=1.0\columnwidth]{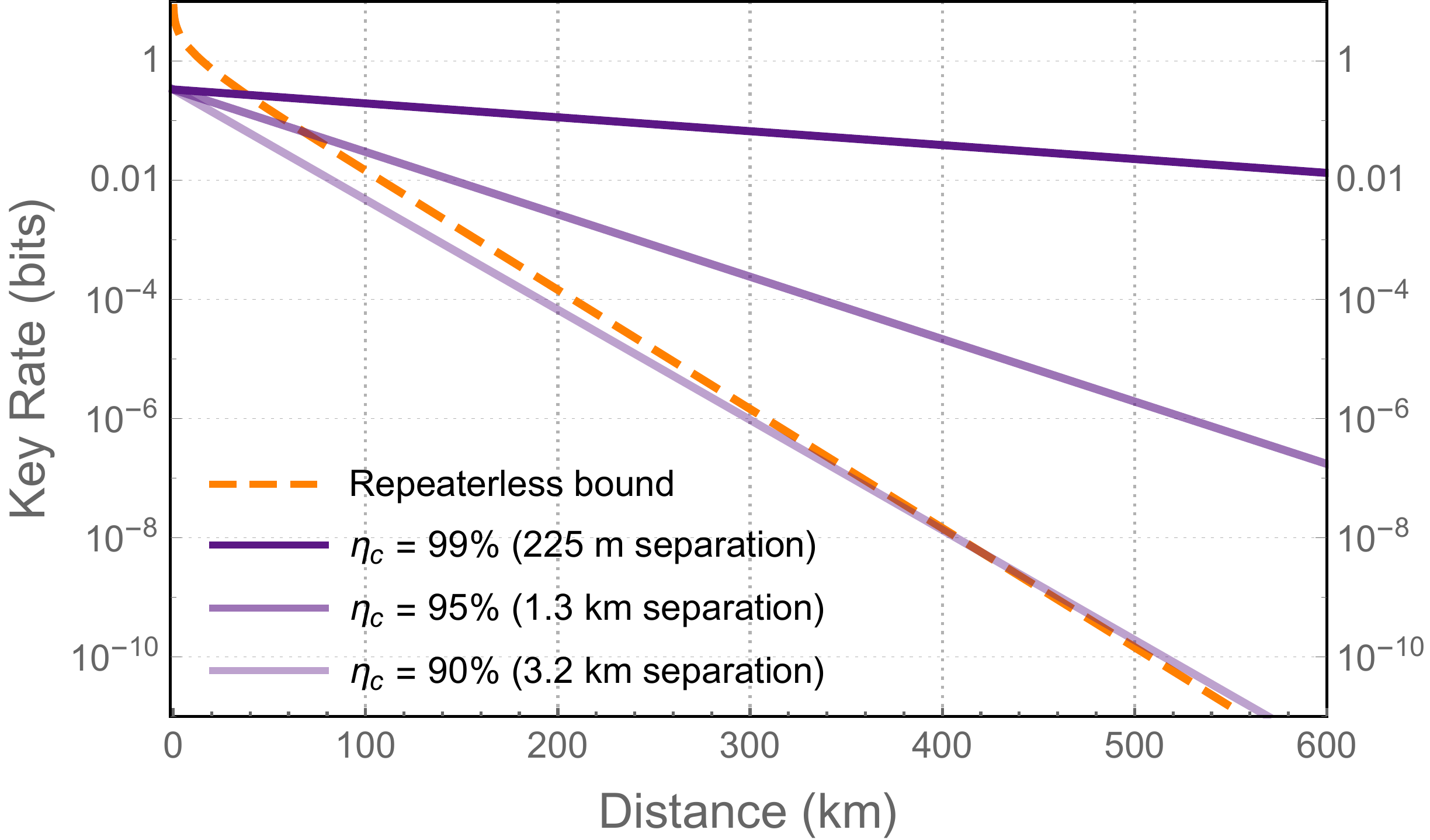}
\caption{\label{example}Example key rates per mode for the three-mode code. For each curve we have optimized the repeater separation (in parenthesis). Note that even with 10\% of coupling loss between fibres and repeaters, the three-mode code can still beat the repeaterless bound starting around the 400 km mark.}
\end{figure}

We can see that unitary one-way quantum repeaters can beat the repeaterless bound asymptotically, which means that the combination of loss and repeater stations effectively turns the lossy bosonic channel into a different channel with lower effective loss, to the point that in a one-way exchange of quantum states we can extract more secret key than one could obtain from a pure-loss channel. In Fig.~\ref{example} we include an example of the key rate over distance obtained with the three-mode code over a range of 600 Km including coupling loss, and compare it to the repeaterless bound.

In this work we have assumed that the repeaters are not introducing any additional noise, which is an unrealistic assumption. We address the limitations due to added noise in our upcoming work \cite{michael}.

This result shows that it is possible to build unsupervised, one-way unitary stations that act as quantum repeaters, which could be placed conveniently every few kilometers along the optical fibre network, allowing us to perform QKD and other distributed quantum information tasks at much increased distances than what we can currently achieve.

Furthermore, some codes show a remarkable resilience to the \emph{additional} coupling loss between fibres and repeater stations, which is encouraging from a technological point of view.

The actual physical realization of a repeater station is a very challenging problem. There are many possible families of bosonic codes and within each family there are many possible choices of specific codes. For each code there is in turn a continuum of Hamiltonians that depend on the initial ancilla state. So there are innumerable Hamiltonians that one could use, but we do not yet understand how to determine if a given Hamiltonian can be implemented (or approximated to a satisfactory degree) by arranging a reasonable number of optical components. Understanding thed connection between a Hamiltonian and its implementation is part of our ongoing work.

\section{Conclusions}
We introduced two families of Hamiltonians that generate a quantum repeater interaction which protects one qubit of information propagating in a lossy bosonic channel. We presented a few examples to show that it is possible to overcome the repeaterless bound even in the presence of a mild coupling loss between the lossy channel and the repeater stations. Our construction is general and it gives the freedom to choose a code space and an initial ancilla state, which translates into the potential to discover particularly convenient Hamiltonians that can be decomposed into a feasible arrangement of a few simple optical elements.

\section{Acknowledgements}
The authors thank Marco Piani, Ryo Namiki and John Jeffers for helpful discussions. This work was supported by NSERC Discovery Grant, Industry Canada and the Army Research Lab.

\bibliographystyle{plainurl}
\bibliography{repeaterBibliography.bib_doi}

\end{document}